\documentclass[sigconf]{acmart}

\usepackage{soul,xcolor}

\AtBeginDocument{%
  }

\setcopyright{acmlicensed}
\copyrightyear{2025}
\acmYear{2025}
\acmDOI{XXXXXXX.XXXXXXX}
\acmConference[SIGIR '25]{Make sure to enter the correct
  conference title from your rights confirmation email}{July 13--18,
  2025}{Padua, Italy}
\acmBooktitle{Proceedings of the 48th International ACM SIGIR Conference on
Research and Development in Information Retrieval (SIGIR '25), July 13--18, 2025, Padua, Italy}
\acmISBN{978-1-4503-XXXX-X/2018/06}
\begin{document}

%%
%% The "title" command has an optional parameter,
%% allowing the author to define a "short title" to be used in page headers.
\title{Multimodal Search in Chemical Documents and Reactions}

%%
%% The "author" command and its associated commands are used to define
%% the authors and their affiliations.
%% Of note is the shared affiliation of the first two authors, and the
%% "authornote" and "authornotemark" commands
%% used to denote shared contribution to the research.
% \author{\textbf{Co-First Authors (?): }Siru, Leo, Ayush, Bryan}
% \authornote{These authors contributed equally to this research.}
% \email{TBD}
% \orcid{TBD}
% \affiliation{TBD}

\author{Ayush Kumar Shah}
\authornote{Authors who led and contributed equally to this research.}
\orcid{0000-0001-6158-7632}
% \email{as1211@rit.edu}

\author{Abhisek Dey}
\authornotemark[1]
% \email{ad4529@rit.edu}
\orcid{0000-0001-5553-8279}
\affiliation{%
  \institution{Rochester Institute of Technology}
  % \city{Rochester}
  % \state{New York}
  \country{}
}
\email{{as1211,ad4529}@rit.edu}

\author{Leo Luo}
\authornotemark[1]
\orcid{0009-0004-0513-1496}
% \email{{mingz5,siruo2,tluo9}@illinois.edu}
\affiliation{%
  \institution{University of Illinois Urbana-Champaign}
  % \city{Champaign}
  % \state{Illinois}
  % \country{USA}
  \country{}
}
\email{tluo9@illinois.edu}

\author{Bryan Amador}
\authornotemark[1]
% \email{ma5339@rit.edu}
\orcid{0009-0007-1398-2751}

\author{Patrick Philippy}
% \email{pmp2516@rit.edu}
\orcid{ 0009-0004-4576-8803}
\affiliation{%
  \institution{Rochester Institute of Technology}
  % \city{Rochester}
  % \state{New York}
  \country{}
}
\email{{ma5339,pmp2516}@rit.edu}

\author{Ming Zhong}
% \email{mingz5@illinois.edu}
\orcid{0000-0001-5728-0224}

\author{Siru Ouyang}
% \email{siruo2@illinois.edu}
\orcid{0009-0004-0513-1496}

% \author{Leo Luo}
% % \email{tluo9@illinois.edu}
% \orcid{0009-0004-0513-1496}
% % \email{{mingz5,siruo2,tluo9}@illinois.edu}

\author{David Mark Friday}
\orcid{0009-0004-3189-5115}

\author{David Bianchi}
% \email{siruo2@illinois.edu}
\orcid{0000-0003-4674-194X}

\author{Nick Jackson}
\orcid{0000-0002-1470-1903}

\affiliation{%
  \institution{University of Illinois Urbana-Champaign}
  % \city{Champaign}
  % \state{Illinois}
  % \country{USA}
  \country{}
}
\email{{mingz5,siruo2,friday1,davidmb2,jacksonn}@illinois.edu}

\author{Richard Zanibbi}
\orcid{0000-0001-5921-9750}
\affiliation{%
  \institution{Rochester Institute of Technology}
  % \city{Rochester}
  % \state{New York}
  \country{}
}
\email{rxzvcs@rit.edu}

\author{Jiawei Han}

\orcid{0000-0002-3629-2696}

\affiliation{%
  \institution{University of Illinois Urbana-Champaign}
  % \city{Champaign}
  % \state{Illinois}
  \country{}
}
\email{hanj@illinois.edu}

%%
%% By default, the full list of authors will be used in the page
%% headers. Often, this list is too long, and will overlap
%% other information printed in the page headers. This command allows
%% the author to define a more concise list
%% of authors' names for this purpose.
\renewcommand{\shortauthors}{Shah et al.}

%%
%% The abstract is a short summary of the work to be presented in the
%% article.

\begin{abstract}
% Efficiently searching and retrieving chemical information from diverse data
% sources is an important challenge in chemical informatics and drug discovery. 
We present a multimodal search tool that facilitates retrieval of
chemical reactions, molecular structures, and associated text
from scientific literature. Queries may combine molecular diagrams,
textual descriptions, and reaction data, allowing users to connect different
representations of chemical information. 
To support this, the indexing process includes
chemical diagram extraction and parsing, extraction of reaction data from text in tabular form, and cross-modal linking of
diagrams and their mentions in text.
We describe the system’s architecture,
key functionalities, and retrieval process, along with expert assessments
of the system. This demo\footnote{\url{https://www.cs.rit.edu/~dprl/reactionminer-demo-landing/}}
highlights the workflow and technical components of the search system.
% Our system is open-source and the demo is available online.
% , contributing towards a
% chemical data retrieval system and analysis.
% \hl{Update link to demo}
\end{abstract}

%%
%% The code below is generated by the tool at http://dl.acm.org/ccs.cfm.
%% Please copy and paste the code instead of the example below.
%%
\begin{CCSXML}
<ccs2012>
   <concept>
       <concept_id>10002951.10003317.10003371.10003381.10003384</concept_id>
       <concept_desc>Information systems~Chemical and biochemical retrieval</concept_desc>
       <concept_significance>500</concept_significance>
       </concept>
 </ccs2012>
\end{CCSXML}

\ccsdesc[500]{Information systems~Chemical and biochemical retrieval}

%%
%% Keywords. The author(s) should pick words that accurately describe
%% the work being presented. Separate the keywords with commas.
\keywords{multi-modal search, reactions, chemical diagrams}
%% A "teaser" image appears between the author and affiliation
%% information and the body of the document, and typically spans the
%% page.
% \begin{teaserfigure}
%   \includegraphics[width=\textwidth]{sampleteaser}
%   \caption{Seattle Mariners at Spring Training, 2010.}
%   \Description{Enjoying the baseball game from the third-base
%   seats. Ichiro Suzuki preparing to bat.}
%   \label{fig:teaser}
% \end{teaserfigure}

\received{18 February 2025}
% \received[revised]{12 March 2009}
% \received[accepted]{5 June 2009}

%%
%% This command processes the author and affiliation and title
%% information and builds the first part of the formatted document.
\maketitle

% COMMENTED
% \hl{The demo paper should answer these questions:}
% \begin{itemize}
%     \item Who is your target user? Why does your system exist and why is it important?
%     \item What problem is your system trying to address? 
%     \item What does your demonstration do and how does it work?
%     \item What does it look like? 
%     \item How does it compare with existing systems? 
%     \item How and when might your technology have an impact, either technically, commercially, or societally?
% \end{itemize}
% % \hl{Find previous SIGIR demo papers with no/less results}

\section{Introduction}
% \hl{Redo the introduction at the end, currently just a rough outline}
% \hl{Introduce the system and motivation}
% \hl{Related systems}
% How does it compare with existing systems? DECIMER, MolScribe group, Pubchem
% What has been indexed

The scientific literature %and patents 
contains vast amounts of chemical knowledge 
represented in textual descriptions and diagrams.
% \hl{
Roughly speaking, molecular and reaction diagrams represent the structure and relation of compounds participating in reactions, while the main text and text labels on reaction diagram entities describe how and why reactions occur (e.g., at what temperature, the yield of a product molecule, pertinent molecular properties, etc.).
As a result, the full story of a reaction is often told using a combination of text and diagrams.
% }

Traditional Chemical Information Retrieval (CIR) systems 
and
commercial platforms such as 
SciFinder\textsuperscript{\textregistered}\footnote{SciFinder\textsuperscript{\textregistered}: \url{https://scifinder.cas.org/}} 
and Reaxys\textsuperscript{\textregistered}\footnote{Reaxys\textsuperscript{\textregistered}: \url{https://www.reaxys.com/}}
% SciFinder\textsuperscript{\textregistered} and
% Reaxys\textsuperscript{\textregistered},
provide extensive text-based and
structural search capabilities. However, these systems 
%often 
do not 
explicitly %fully
associate %integrate 
molecular figures with their textual descriptions, making it difficult
to retrieve %entire 
reactions found in diagrams %records 
with their details and context provided in text.
%\hl{such as XXX, YYY, ZZZ} primarily rely on
%text-based indexing, which limits their ability to connect structured chemical
%representations with their corresponding textual context. 
Existing systems 
%These earlier
%systems 
are also designed to return individual compounds or full documents as results, rather than returning text passages where reaction descriptions are found. 
These limitations present
challenges for chemists, patent examiners, and researchers seeking to retrieve
relevant compounds, reactions, or synthesis protocols along with their contexts efficiently. 
% Existing
% commercial platforms, such as 
% SciFinder\textsuperscript{\textregistered}\footnote{SciFinder\textsuperscript{\textregistered}: \url{https://scifinder.cas.org/}} 
% and Reaxys,\textsuperscript{\textregistered}\footnote{Reaxys\textsuperscript{\textregistered}: \url{https://www.reaxys.com/}}
% % SciFinder\textsuperscript{\textregistered} and
% % Reaxys\textsuperscript{\textregistered},
% provide extensive text-based and
% structural search capabilities. However, these systems often do not fully
% integrate molecular figures with their textual descriptions, making it difficult
% to retrieve entire reaction records with their contextual details.

To address these challenges, 
our system supports direct retrieval of relevant passages, 
which are returned along with their associated molecular structures.
This includes structures extracted from
molecular diagrams and referenced in the text by common name (e.g., `chromene'), IUPAC name \cite{Skonieczny2006},
% \footnote{\url{https://iupac.org/what-we-do/nomenclature} \hl{RZ: Is there a better link/reference?}}
or figure identifier (e.g., molecule `34', or molecule `4b').
Automatically extracted reaction records are also generated from passages using \emph{ReactionMiner} \cite{zhong-etal-2023-reaction}, and provided alongside passages and their associated compounds. The reaction records enable researchers to explore linked reaction steps.
%within
%scientific documents.
 %This improves 
%accessibility to structured chemical information in scientific literature.
% More recent approaches, such as
% Text2Mol~\cite{Edwards2021} and MoleculeSTM~\cite{moleculeSTM2022}, leverage
% deep learning approaches, including GNN and multimodal transformers to generate
% molecule-text representations but are not designed
% for traditional passage-level search and retrieval. Our system differs in that
% it provides passage-level retrieval rather than document-level retrieval and
% directly links extracted molecular diagrams with text mentions and reactions, improving
% accessibility to chemical and reaction-related information.
%Our system addresses this gap by indexing text passages, molecular structures,
%and chemical reactions from scientific documents, enabling 
Passages may be searched using text queries, molecular structure
queries in SMILES (Simplified Molecular Input Line Entry System) strings \cite{Weininger1988}, or a combination of the two.
SMILES is frequently used in chemoinformatics, and can be readily generated by a number of commonly used drawing tools (e.g., ChemDraw or Marvin). 
This multimodal search model facilitates direct navigation
between reaction text, associated molecule diagrams, and extracted chemical
entities. 
The system supports text-based search using BM25, and SMILES-based molecular search using structural similarity and substructure matching provided by RDKit\footnote{\url{https://www.rdkit.org}}.
% REDUNDANT
%and multimodal
%search combining textual and structural queries to retrieve reactions involving
%specific substructures along with their contextual descriptions.

% REMOVE? Contributions are stated in 3rd paragraph.
%Our system introduces a unified retrieval approach that integrates chemical text
%mining, molecular diagram parsing, and reaction extraction, linking chemical
%figures with their respective textual mentions. The key contributions include
%extraction and indexing of compounds and reactions from both text and molecular
%diagrams, passage-level linking of molecular diagrams with textual reaction
%descriptions, and multimodal search that allows users to search via text,
%molecular structure, or a combination of both. Additionally, direct navigation
%of reaction records enables researchers to explore linked reaction steps within
%scientific documents.

% \hl{RZ: This paragraph may need to be slightly updated to link with revisions above.}
\textbf{Related Works.} Early CIR systems such as
ChemXSeer~\cite{chemxseer} focused on extracting and
indexing chemical names from tables in PDFs, in order to allow users to search
by molecule name or formula. Later efforts, such as
TREC-CHEM~\cite{trecchem2009}, introduced the concept of document-level
retrieval for chemistry-specific tasks, curating a dataset of patents and
employing manual relevance assessments to evaluate retrieval effectiveness.
Recent advances in deep learning have focused on cross-modal learning to align
structured molecular representations with textual descriptions. Both
Text2Mol~\cite{Edwards2021} and
MoleculeSTM~\cite{liuMultimodalMoleculeStructure2023a} adopt joint learning
approaches that embed chemical structures and text into a shared embedding
space, facilitating retrieval across modalities. Text2Mol employs graph neural
networks (GNNs) \cite{scarselliGraphNeuralNetwork2009} and perceptron models to
predict the most appropriate SMILES representation given a textual query,
addressing the task of molecule retrieval from natural language descriptions.
Similarly, MoleculeSTM is designed for structure-text retrieval, retrieving
chemical structure from textual descriptions and vice versa by leveraging a
multimodal transformer within a contrastive learning framework.
% to align
% molecular structures with text-based knowledge. 
However, these systems focused
on document-level indexing rather than passage-level search, whereas our system
enables structured retrieval by directly linking extracted molecular diagrams,
textual mentions, and reaction descriptions.
OpenChemIE~\cite{fanOpenChemIEInformationExtraction2024a} extracts reaction data
from text, tables, and figures using modality-specific models, similar to our
approach of combining text-based reaction extraction and molecular diagram
parsing. However, unlike OpenChemIE, our system
integrates extraction with passage-level retrieval.
% , enabling search and
% cross-modal linking of molecular structures, textual mentions, and reactions.

% enabling zero-shot generalization to tasks such as generating textual descriptions for molecules and modifying molecules based on text instructions.
% Recent approaches such as Text2Mol~\cite{Edwards2021} and
% MoleculeSTM~\cite{liuMultimodalMoleculeStructure2023a} leverage deep learning techniques to jointly learn and integrate two modalities:
% % \hl{RZ: map in exactly which way? Generating textual descriptions for molecules in diagrams? Through linking? Clarify.} 
% chemical structures and textual descriptions through a common aligned embedding space, learning through the paired set of molecules and text descriptions.
% \hl{(RZ: What is the indexing and/or retrieval purpose of systems before TREC-CHEM? Relationship with points in first two paragraphs?)}
% Text2Mol \cite{Edwards2021} uses GNNs \cite{scarselliGraphNeuralNetwork2009} and
% perceptron models to predict the most appropriate SMILES representation given a
% user-provided molecular description, while MoleculeSTM \cite{liuMultimodalMoleculeStructure2023a} employs a multimodal
% transformer network with contrastive learning to select the most likely text
% description from a set of options for a given SMILES input. 

% TREC-CHEM~\cite{trecchem2009} was one of the first shared tasks for
% evaluating chemistry-specific retrieval, using a manually curating patent dataset for
% document-level retrieval. 

Our demonstration showcases how structured indexing and linking of chemical information
across text and figures can improve chemical information retrieval and provide a
more comprehensive and flexible search framework for chemists and researchers.

\section{Indexing Extracted Compounds and Reactions}

In our system, textual and graphical content are processed through two parallel
pipelines, whose outputs are later used to build a unified index of chemical
entities, paragraph texts, and reactions.

\begin{figure}[!tb]
  \centering
  \includegraphics[width=\linewidth]{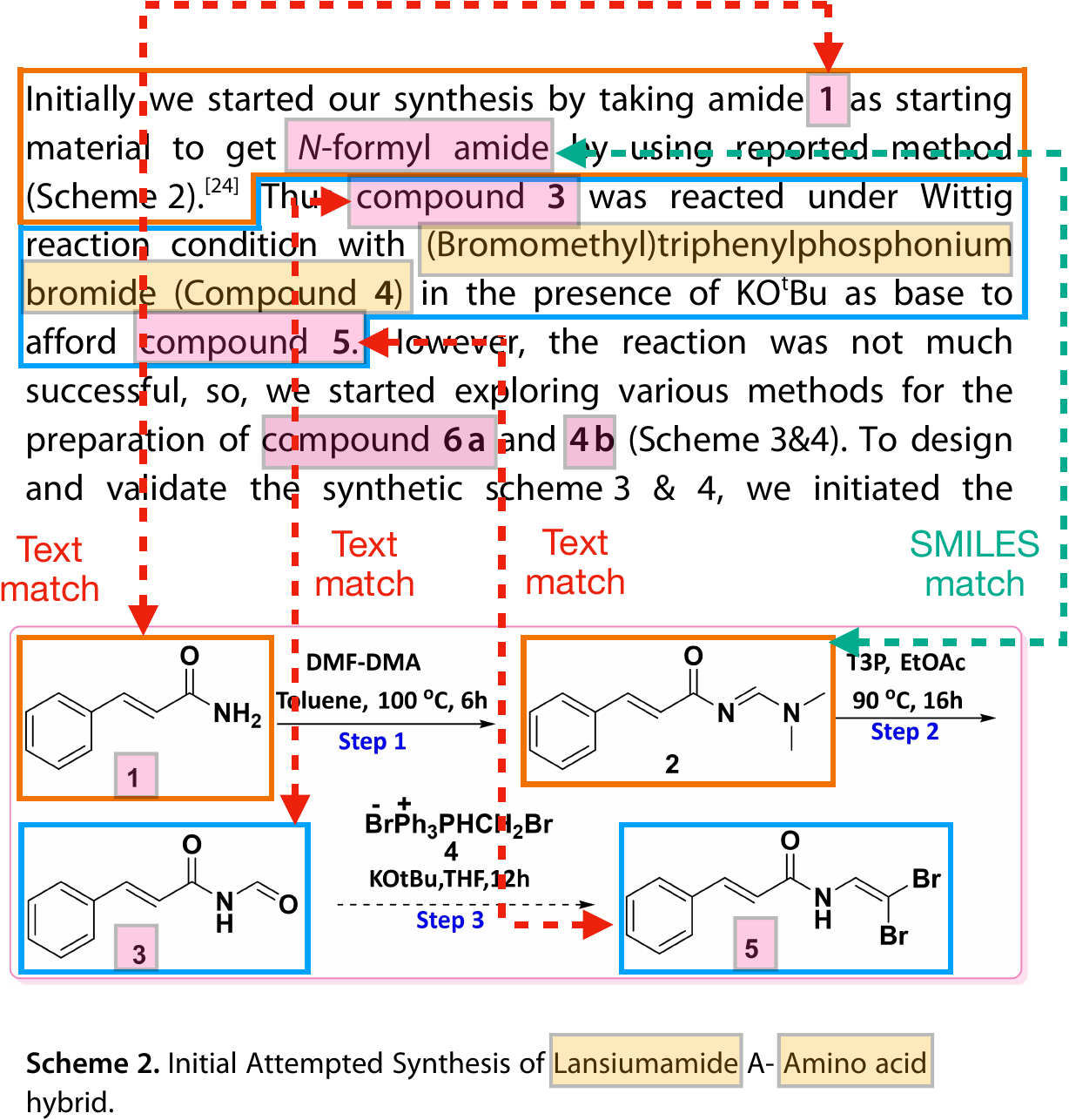}
  % \caption{Linking reaction text and molecule diagrams}

\caption{\textbf{Text/diagram compound extraction and compound--passage linking.} 
Two passage types are shown: (1) reaction passages from 
\emph{ReactionMiner} (2 boxes) and (2) a single text passage 
containing both extracted compounds.
%s extracted using PyTesseract. 
Highlighted text denotes extracted chemical entities: pink for mentions in
molecular diagrams, yellow for unmatched mentions. Matches come from
(1) \emph{Text matching} via Levenshtein distance and (2) \emph{SMILES 
matching} via \emph{Tanimoto Similarity}. Highlight colors (e.g., orange and blue) 
indicate molecules \& reaction text linked to the same reaction passage.}

  % \caption{Extracting compounds from text and diagrams, and
  %   compound--passage linking.
  % Illustration showing two categories of passages: those identified by
  % \emph{ReactionMiner} (boxed) and other unboxed text blocks. In the boxed regions,
  % colored highlights mark token‐level extractions for reactants, products, and
  % other chemical entities: pink
  % highlights refer to text mentions matched to a molecule diagram in the 
  % document, while yellow highlights refer to text mentions that do not link to a
  % specific diagram. 
  % % while orange or blue highlights group reactants/products linked to the
  % % same reaction. 
  % These matches are formed through two methods:
  % (1)~Levenshtein‐based text matching (e.g., `compound~5' $\leftrightarrow$ diagram
  % labeled~`5'), and (2)~SMILES‐based matching, where each text token (via
  % ChemDataExtractor + OPSIN) and figure diagram (via \emph{ChemScraper}) is converted to
  % SMILES for Tanimoto‐based similarity checks. Other compound mentions (in
  % yellow) in both unboxed and boxed passages
  % are extracted but do not link to a specific diagram; they are still converted
  % to SMILES for indexing.
  % Multiple molecule diagrams referencing the same reaction share consistent
  % color highlights (e.g., orange for `amide 1' and `N-formyl amide' along with
  % their reaction text, and blue for `compound 3' and `compound 5' with their
  % corresponding reaction text), indicating they belong to the same
  % reaction block.}
  \Description{Extraction and linking figure}
  \label{fig:linking}
\end{figure}

\textbf{Text mining for extracting reaction information.}
We employ 
% a reaction‐extraction, 
\emph{ReactionMiner \cite{zhong-etal-2023-reaction}}, a
pipeline that processes text extracted from PDF documents to isolate
and categorize reaction‐related content. 
First, the text is segmented into
reaction-related sentences through product-indicative keywords and topic
modeling \cite{choi-2000-advances} for defining the contextual boundary. A
large language model, LLaMA3.1-8b
\cite{grattafioriLlama3Herd2024}, fine-tuned with LoRA \cite{hu2021lora}, is then used to identify relevant chemical entities such as reactants, products, and catalysts, along with key conditions (e.g., temperature, reagents, or solvents).
% to locate references to
% reactants, products, catalysts, 
% and relevant condition details from the document text. 
Each reaction mention is 
associated with segmented text bounding boxes in the PDF (see
Figure~\ref{fig:linking}), enabling direct navigation to the 
underlying paragraph. By grouping identified mentions into coherent reaction 
units, a structured record for each reaction step present in the 
text is established.

\textbf{Extracting SMILES from document text.}
In addition to the reaction records extracted by \emph{ReactionMiner}, we also extract individual compounds from %other text regions in t
document text. %to detect compounds that might not be
%for individual compounds.
%explicitly part of a reaction. For these passages, 
These additional SMILES annotations ensure that compounds mentioned both inside and outside of
reaction descriptions can be retrieved through the search interface.
For indexing compounds in text, we first use PyTesseract to convert document page images to 
text, which is then passed to 
\emph{ChemDataExtractor2.0's} \cite{mavracic_2021} Chemical Named Entity Recognition (CNER) system to identify molecule names.
%chemical
%molecule phrases. 
Each recognized name is converted into a SMILES string via
\emph{OPSIN} \cite{opsin},
and any passage containing at least one valid SMILES is retained for indexing.

\textbf{Molecular diagram extraction and parsing.}
We use YOLOv8, an improved version of scaled
YOLOv4~\cite{wang_scaled-yolov4_2021},
% with enhanced performance and efficiency,
for detecting molecular regions in documents. 
We then employ \emph{ChemScraper}~\cite{shah_chemscraper_2024} to parse molecular
diagrams from detected PDF regions (see Figure~\ref{fig:linking}) through two complementary pipelines: a
\textit{born‐digital} approach for vector images representing characters and
geometric objects, and a \textit{visual parsing} \cite{shahLineofSightGraphAttention2023}
approach for pixel-based raster images. 
In the born‐digital parser, \textit{SymbolScraper}
\cite{shahMathFormulaExtraction2021} accesses low‐level PDF drawing commands to extract lines,
polygons, and characters directly from the PDF. The visual parser works on
raster images, applying the \textit{Line Segment Detector (LSD)}~\cite{gioiLSDLineSegment2012}
and watershed algorithm to detect line primitives and text regions. Together,
these methods yield a set of graphical elements (e.g., atoms, bond lines, named
functional groups) and their local connections.

\emph{ChemScraper} then constructs a \textit{visual structure graph} 
using a Minimum Spanning Tree (MST) and rewrite rules for
born‐digital diagrams and a segmentation‐aware, multi‐task neural network for
raster images. 
The visual graph is then converted to a molecular graph, where bonds become
edges and implicit carbon atoms are inferred from line intersections.
% Bonds are labeled (single, wedge, double, etc.), and lines at
% intersections are treated as implicit carbon atoms. The resulting graph is
% converted into a full molecular graph, 
The final molecular graph is stored in \textit{CDXML format}
to retain both chemical and visual structure. This format is 
converted into SMILES for indexing.

% Talk about chemscraper \cite{shah_chemscraper_2024}, reactionminer \cite{zhong-etal-2023-reaction} and linking

{\textbf{Compound-Passage Linking and Multimodal Indexing.}
Once text‐based reaction information and text and diagram‐based molecular structures are
extracted, they must be unified into a cohesive representation that supports
flexible querying. 
%The following steps outline how these components are linked
%and indexed to enable multimodal retrieval.
There are two types of passages in our system: (1) those extracted by
\emph{ReactionMiner} (boxed text in Figure~\ref{fig:linking}), and (2) those
extracted from general text regions using PyTesseract (unboxed). 
% \hl{RZ: Do we not link all detected mentions in text with diagrams as well? I'm not sure this is clear here. -- AS: We link the detected mentions of reactants and products from reaction miner texts to diagrams only. For texts which are not from reaction miner, we only convert the detected mentions in text to SMILES, and these are available in index, but not linked to any diagrams -- This is the difference compared to Abhisek's system in terms of diagram linking.}

For passages extracted by \emph{ReactionMiner}, relevant text fields include
reactants, products, catalysts, and yields. 
% \hl{@Bryan - check the matching strategy described below: specifically add missing clarity on using text vs SMILES based matching, which is used by default, and which one is used when, also mention any thresholds involved.}
We focus on linking the reactants and products in reaction records 
with their corresponding molecular diagrams using two approaches:
\textit{1. Token-based text matching:} Text mentions of diagram labels for reactants and products
are identified using regular expressions,
%to clean and split them into meaningful tokens
%(e.g., removing unnecessary parentheses, extra words like "compound," and
%separating multiple compound names). 
%Resulting tokens are 
and then matched with the
nearest diagram text label by a normalized Levenshtein similarity ratio\footnote{\url{https://rapidfuzz.github.io/Levenshtein/levenshtein.html\#ratio}}, ensuring minor variations in naming
do not prevent linkage. An example is shown in red in
Figure~\ref{fig:linking}, where `compound 5' is matched with the
diagram labeled `5.'
\textit{2. SMILES-based fingerprint matching:} 
% In cases where text-based matching
% is insufficient, 
% We convert compound names to SMILES.
%and diagram
%representation into a common format—SMILES. 
The text tokens are first processed
through \emph{ChemDataExtractor2.0} \cite{mavracic_2021} and \emph{OPSIN}
\cite{opsin}, following the same
approach described earlier for non-reaction text, while SMILES from diagrams
are extracted by \emph{ChemScraper} \cite{shah_chemscraper_2024}. Each SMILES representation
undergoes molecular fingerprinting, producing a binary vector encoding the
molecular graph's connectivity patterns. 
To determine similarity, we compute the \textit{Tanimoto Similarity}
\cite{finger_search} %between pairs of fingerprints. The Tanimoto coefficient
measuring the overlap between fingerprint vectors, giving a similarity score
between 0 and 1. 
The diagram with the highest \emph{Tanimoto Similarity} score with a passage SMILES 
% and a diagram SMILES surpasses a threshold value, they 
are linked. An example is shown in blue in
Figure~\ref{fig:linking}, where `N-formyl amide' is matched with the
diagram labeled `2.'
% \hl{Check with Bryan.} \hl{RZ: Provide the threshold value.}
% Otherwise, the system falls back to token-based text matching to ensure robust
% compound linking across text and figures. 
% \hl{
If a compound can be linked by both strategies, the strategy with the higher score between the
normalized \emph{Levenshtein similarity ratio} and the \emph{Tanimoto Similarity} is chosen.
% }
% (anything with similarity of 0 is dropped).}
% \hl{RZ: The 'fall back' is unclear here -- what type of match is expected from the chemical name if the SMILES fails, for example?: @Bryan could you check this? - maybe after revising based on previous comment solves this. \textit{Bryan}: There isn't a threshold value. Linking with SMILES is done wherever is possible and with string wherever is possible; if a compound can be linked by both, the strategy with highest confidence is chosen (anything with similarity of 0 is dropped).}

In the final index, passages are annotated
with associated SMILES and reaction entities. This enables retrieval of
specific compounds as well as reaction-related information, linking molecular
structures to their reaction context.
During indexing, IUPAC names in text are also tokenized into constituent groups
\cite{Rajan2021} to improve retrieval. For example, 
% ...(\hl{RZ: Add example})
an IUPAC name, `N-((E)-2-bromo-2-phenylvinyl)-cinnamamide'
would be tokenized as `N E 2 bromo 2 phenyl vinyl cinnamamide'.

\section{Multimodal Search: Suzuki Coupling Papers}
Our system supports three primary search models based on different query types:
text search, molecular structure/substructure search (from SMILES), and multimodal search
combining text and molecular queries. 
Text search is handled using BM25 \cite{Robertson1994OkapiAT} as implemented in PyTerrier \cite{pyterrier2O21}, %which ranks passages based on
%textual relevance, 
allowing users to search for compounds, reaction conditions,
or chemical properties in text. %using natural language queries. 
Molecular structure
search is %is based on cheminformatics techniques provided by
provided by RDKit, enabling both
exact molecule matching and substructure searching.
%using SMILES-based
%fingerprinting. 
Multimodal search integrates results from both models,
%performing fusion and 
by re-ranking results to prioritize passages that contain both
textual and molecular matches.
% ensuring more comprehensive retrieval of relevant chemical information.
% These search methods allow retrieval of molecular and reaction-based
% information, whether the entities appear in text descriptions or are represented
% as molecular diagrams.

% \textbf{Passage and SMILES Indexing.} 
% \hl{Bryan - check these numbers. \textit{Bryan}: Corrected the number of papers}. 
For this demonstration, we indexed seven
research papers and six supplementary
information documents related to Suzuki coupling reactions, provided by  chemists at the University of Illinois. 
% \hl{Abhisek/Bryan - provide/check these numbers.} 
The dataset includes a total of 1282 extracted passages, out of which 538 are indexed (passages without links to a reaction or compound name are removed), 383 unique SMILES string, and 219 extracted reactions.
These passages are linked to diagrams and SMILES as described in the previous section.
% but passages without links to a reaction or compound name are not indexed.
%Each passage in the index is linked to
%individual SMILES or a set of SMILES extracted from text (both
%\emph{ReactionMiner} and other passages) and diagrams linked to reactants and
%products identified by \emph{ReactionMiner}. 
%These
Molecule SMILES are also searchable within detected chemical entities in passages and reactions. 
% \hl{RZ: Is this previous statement correct? -- Yes}
%forming a structured index. 
This structured linking allows users to retrieve
molecular and reaction information, whether entities
appear in text descriptions or are represented as molecular diagrams.
%of
%reactants and products.
% To
% maintain efficiency, duplicate entries are avoided—each passage and SMILES
% representation is stored uniquely, even if multiple mentions exist within the
% same passage.

\begin{figure}[!tb]
  \centering
  \includegraphics[width=0.8\linewidth]{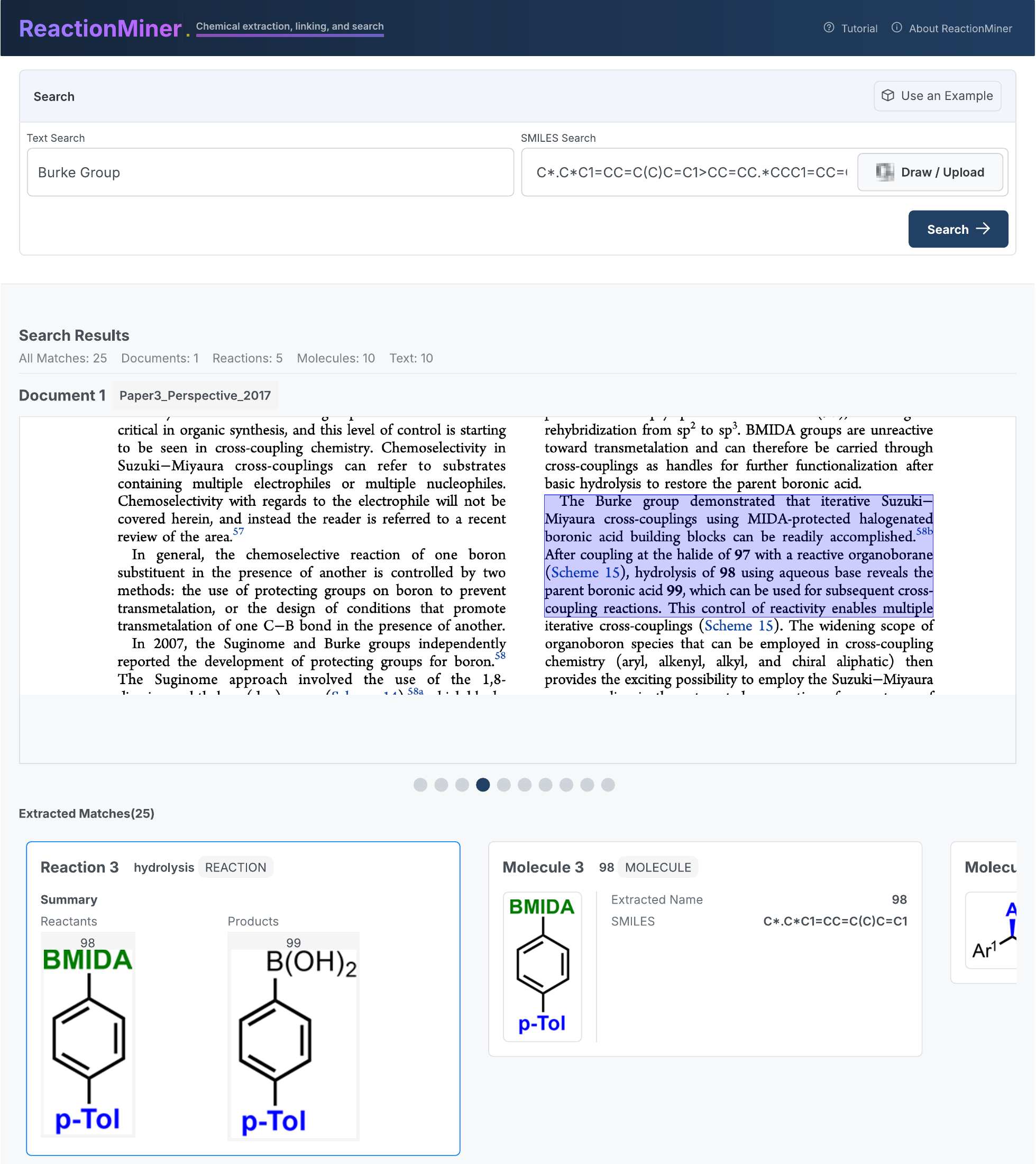}
\caption{Multi-modal search results for a text and Reaction SMARTS query.
Results are organized by document, with matched passages linked to extracted
reactions, molecular structures, and highlighted text mentions. Key reaction
details, including reactant (`98') and product (`99') in both text and diagrams,
along with their predicted SMILES representations, are displayed. Users can
navigate directly to relevant sections within each document, with highlighted
passages indicating the corresponding matches.}
%   \caption{Multi-modal search results for a text + reaction SMARTS query.
%   Results are grouped by document, with matched passages linked to extracted
% reactions, molecules, and text highlights. Key reaction details, including
% reactant (`98') and product (`99') in both text and diagrams, as well as
% corresponding predicted SMILES are displayed,
% with direct navigation to relevant sections, highlighted within each document.}
  \Description{Search results interface showing reaction, molecule, and text
  matches for a multi-modal query, grouped by documents with highlighted chemical entities.}
  \label{fig:search_results}
\end{figure}

% \textbf{Text-Only Search.} For text-based queries, passage retrieval is
% performed using the BM25 \cite{Robertson1994OkapiAT} ranking model, which scores
% relevance based on term
% frequency and document length normalization. This allows users to search for
% multiple compounds, reaction conditions, or chemical properties using natural
% language queries. The retrieved passages may contain references to chemical
% reactions, molecular descriptions, or experimental setups. Additionally,
% passages identified by \emph{ReactionMiner} are annotated with structured
% information such as reaction conditions, catalysts, and temperature, providing
% additional details to users within an expandable view.
% The text retrieval pipeline also applies domain-specific tokenization by
% breaking down IUPAC names into smaller constituent groups to improve passage
% ranking.

% \hl{RZ: Removed text search, it repeated earlier content.}

\textbf{Molecular SMILES and SMARTS Search.}
%For molecular structure queries using SMILES, retrieval is performed using
%RDKit's molecular
%similarity search methods. The input 
Query SMILES and candidate SMILES in the index are converted into fingerprint
vector representations using Morgan Fingerprinting (2048 bits), and 
% candidate SMILES in the index 
are ranked using \emph{Tanimoto Similarity}, described earlier.
The system also accommodates structured queries specifying entire chemical
reactions using \emph{Reaction SMARTS} (SMILES Arbitrary Targets Specification)
notation. This format extends SMILES strings to include `>' separating
reactants, reagents, and products, while `.' separates individual SMILES
compounds within each category, as seen in Figure~\ref{fig:search_results}. To
process these queries, SMILES compounds are
extracted from the SMARTS string, and each is
%the Reaction SMARTS string is reformatted by
%replacing `>' and `.' with spaces, effectively breaking it into separate
%molecular queries. Each compound in the reaction query is 
matched against
indexed molecular structures and reaction passages.

This allows retrieval of passages describing reactions described in SMARTS,
while also providing access to details provide in 
\emph{ReactionMiner} records linked to passages. Information such as reactants,
products, catalysts, reaction conditions, and temperature are presented 
%in an
%expandable chip in the 
for retrieved passages. This enables users to search for
entire reactions rather than just isolated compounds, improving retrieval of
contextually relevant reaction information.

\textbf{Multimodal Search.} 
For multimodal queries, containing both text and SMILES (or Reaction SMARTS), as
shown in Figure~\ref{fig:search_results},
the initial SMILES candidates are obtained using \emph{Sub-structure search}. We
find that including all possible sub-structures as valid candidates for
re-ranking text candidates leads to better performance as \emph{Tanimoto
Similarity} can be unpredictable for very specific information needs when
combined with text.
%ensuring that structurally similar compounds are retrieved.
% \hl{RZ: It is unclear how the user selects simliarity or exact sub-matching; are we not using both with thresholding? Are we repeating ourselves here? - @Abhisek -- Could you clairfy on this?} \hlc{We use tanimoto for SMILES/SMARTS only queries, sub-structure for multimodal. Clarified above}
If multiple SMILES are
provided, the retrieved passages for each individual SMILES query are
aggregated. 
%The system evaluates the overlap between text-based and SMILES-based
%passage retrieval, and 
Results from
text-based and SMILES-based retrieval are combined. A fusion step adjusts rankings to prioritize passages
containing a higher number of matched SMILES.
% \textbf{Fusion and Ranking.}
%To integrate multimodal search results, 
Retrieved passages are re-ranked based
on their BM25 text relevance score and the presence of SMILES matches. 

\textbf{Reaction Navigation.} The system also provides a dedicated
reaction navigation feature for each retrieved document. When a user selects a
passage, they can inspect all extracted reactions from the associated document in a
structured list, with each reaction entry pointing to the relevant PDF page and
bounding‐box highlights, as shown in Figure~\ref{fig:search_results}. This
approach lets users explore multiple reaction
mentions in context, making it easier to follow complex procedures, compare
alternative synthetic routes, or identify recurring reagents and intermediates
within a single publication.
% \hl{RZ: CHeck whether this document view will be available in the demo. -- AKS: Probably not, might remove this if that's the case. RZ: I would comment it out and not delete it, just in case.}

% RZ: Removed for space, add to conclusion/in expert evaluation.
% This indexing and retrieval strategy allows users to flexibly search using
% natural language queries, chemical structures, or full reaction specifications.
% By integrating structured chemical knowledge from both text and molecular
% representations with reaction navigation, the system enables exploration of
% compound mentions, reaction
% mechanisms, and experimental conditions in a unified manner.
\section{Expert Evaluation by Chemists}
To evaluate the system’s effectiveness for chemists, we conducted an expert
assessment with researchers at the University of Illinois. The system
effectively retrieved relevant chemical information, linking molecular diagrams
and text-based reaction details to chemical names or SMILES queries. 
For example, as shown in Figure~\ref{fig:search_results}, a multi-modal search with
the text query `Burke group' and a Reaction SMARTS string 
successfully retrieved passages with relevant reactions matching the SMILES
query, which were associated with the `Burke group', as shown in the highlighted
text and corresponding reaction diagrams displayed below the document image.
Chemists found the ability to
click on molecule `cards’ and navigate directly to the corresponding section in
the document particularly useful. The structured reaction output captured key
experimental details such as yield, catalysts, solvents, and temperature,
enhancing the accessibility of reaction data. 
% Additionally, the structured reaction output captured key reaction details
% such as yield, catalysts, solvents, and temperature, making it easier to extract
% relevant reaction data. 
The reaction and molecule cards serve as a structured extractive summary
of the paper, while also providing navigation links to their original context.
Note that the `Reaction 3' in
Figure~\ref{fig:search_results} does not show
these additional details as they were not available in the text. 
% A particularly
Notably, the system retrieved
derivatives of a queried molecule, such as `benzo[b]thiophen-2-ylboronic acid,’
which was relevant to the SMILES query `C1=CC=C2C(=C1)C3=CC=CC=C3S2'
(dibenzothiophene) but not explicitly searched for. Overall, the chemists were
able to find the information they were hoping for, and the retrieved results
were useful for their research.

% Despite these strengths, some areas require improvement. 
While reaction details
were generally well extracted, experts recommended incorporating additional
metadata such as `equivalents' and mol\% of catalysts, which are essential for
exporting data to electronic lab notebooks. When using a combination of text and
SMILES queries, users found it challenging to determine whether retrieved
results were more influenced by text-based or structure-based matching,
suggesting a need for greater transparency in the effect of text vs. SMILES on
ranking. Additionally,
filtering options to view reactions, molecules, and text separately would
improve usability, allowing chemists to focus on the most relevant data. Another
key area for improvement is enhanced diagram-text linking, as some extracted
text mentions were not associated with their corresponding molecular diagrams.
Addressing these issues would further enhance the system’s utility for chemical
research.
% and literature exploration.

\section{Conclusion and Future Work}
This work presents a multimodal search system that integrates text and molecular
structure retrieval, enabling passage-level search with structured linking
between chemical entities, molecular diagrams, and reaction descriptions. By
combining BM25 for text, RDKit-based molecular similarity search, and a fusion
mechanism for multimodal queries, our system improves access to chemical
knowledge in scientific literature.
The expert evaluation with chemists demonstrated the system’s usability, with
researchers successfully retrieving relevant chemical information, including
useful molecular derivatives and structured reaction details.

Future work will focus on enhancing retrieval effectiveness with dense
embeddings and cross-modal search, leveraging transformer-based models to
improve ranking and semantic matching across text and molecular representations.
While the system currently matches chemical names to diagrams via SMILES
translation, inspired by Text2Mol~\cite{Edwards2021}, we aim to explore query
expansion within an aligned multimodal embedding space. This approach would
expand text queries to incorporate corresponding molecular diagrams or SMILES
representations and extend SMILES queries to include relevant text-based
descriptors, improving retrieval flexibility.
% We can match common and IUPAC names to diagrams through SMILES translation, but 
% inspired by Text2Mol~\cite{Edwards2021}, we aim to explore query expansion
% within an aligned multimodal embedding space, expanding text queries to include 
% corresponding molecular diagrams or SMILES, and SMILES queries to include
% corresponding text queries for search.
Additional directions include scaling the system to index larger collections and
integrating external chemical databases, and refine filtering mechanisms to
improve user experience.
A more mature version of our system may find use in 
chemical research, industry, and patent analysis, reducing time spent on
literature review and supporting efficient retrieval of structured chemical information. 
% By bridging textual and molecular
% , and incorporating interactive query refinement are also key
% directions.

% The system has applications in chemical research, industry, and patent
% analysis, reducing time spent on literature review and supporting efficient
% retrieval of structured chemical information. By bridging textual and molecular
% representations, it provides a more flexible and context-aware search tool for
% chemists, researchers, and professionals in drug discovery and material science.
% \hl{RZ: incorporate multimodal and dense embddings + cross-modal search
% mentioned in Paragraph 4 of the intro here.}
% \hl{Highlight information needs}
% \hl{Briefly address this question from SIGIR here: "How and when might your
% technology have an impact, either technically, commercially, or societally?"}

\begin{acks}
This work was supported by the National Science Foundation USA (Grant
\#2019897, Molecule Maker Lab Institute). We also thank Matt
Berry, Kate Arneson, Bingji Gao, Sara Lambert, and other members from the NCSA team who helped create the
online system.
\end{acks}

%%
%% The next two lines define the bibliography style to be used, and
%% the bibliography file.
\bibliographystyle{ACM-Reference-Format}
\bibliography{sample-base,references}

\end{document}